\documentclass[12pt]{article}

\begin{document}

\newcommand{\beq}{\begin{equation}}
\newcommand{\eeq}{\end{equation}}
\newcommand{\bea}{\begin{eqnarray}}
\newcommand{\eea}{\end{eqnarray}}
\newcommand{\eL}{{\cal L}}
\newcommand{\half}{\frac{1}{2}}
\newcommand{\J}{\bf J}
\newcommand{\bP}{\bf P}
\newcommand{\G}{\bf G}
\newcommand{\K}{\bf K}
\newcommand{\M}{{\cal M}}
\newcommand{\bu}{\bf u}
\newcommand{\la}{\lambda}

\begin{flushright}
physics/9803029\\
DOE-ER-40757-110\\
UTEXAS-HEP-98-3
\end{flushright}

\hfil\hfil

\begin{center}
{\Large {\bf $SU(3)$ Revisited}}
\end{center}
\begin{center}
\today
\end{center}
\hfil\break
\begin{center}
{\bf Mark Byrd \footnote{mbyrd@physics.utexas.edu} and E. C. G. Sudarshan\\}
\hfil\break
{\it Center for Particle Physics \\
University of Texas at Austin \\
Austin, Texas 78712-1081}
\end{center}
\hfil\break

\begin{abstract}

The ``$D$'' matrices for all states of the two fundamental representations and octet are shown in the generalized Euler angle parameterization.  The raising and lowering operators are given in terms of linear combinations of the left invariant vector fields of the group manifold in this parameterization.  Using these differential operators the highest weight state of an arbitrary irreducible representation is found and a description of the calculation of Clebsch-Gordon coefficients is given.

\end{abstract}

\section{Introduction}

In our understanding of particle physics, studying the group SU(3) has helped tremendously.  It has given us an organization to the plethora of ``elementary'' particles through the Eightfold way\cite{efw} and then led to the quark description of hadrons \cite{qm}.  This, in turn, led to the fundamental theory of the strong nuclear interactions known as the color SU(3) of the now widely accepted standard model \cite{w}.  It has also had numerous successes in phenomenological models such as the nuclear SU(3) model of Elliot \cite{e}, and the Skyrme-Witten model \cite{sm}.  Its algebra has been utilized extensively for these applications but its manifold has not.  In most cases, due to the intimate relationship between the algebra of a Lie group and the group itself (subalgebras correspond to subgroups etc.), this description has been enough.  Also, since the group manifold of SU(3) is 8 dimensional, it is not prone to ``visual'' analysis.  Recently however, the manifold has been used for the study of quantum 3 level systems and geometric phases \cite{m1}, \cite{m2}.  The subgroups and coset spaces of SU(3) are listed in \cite{m2} along with a discussion of the geometry of the group manifold which are relevant to the understanding of the geometric phase.  It should therefore be no surprise if the group and group manifold lead to further understanding of physical phenomena beyond what the algebra has already accomplished.  Further study of its structure may very well lead to an even greater understanding of nature and the way its symmetries are manifest. 

Here, the raising and lowering operators of the group are given in terms of differential operators.  The states of the fundamental representations are given in terms of the Euler angle parameterization.  A highest weight state is given for all irreps (irreducible representations) that will enable the calculation of any state within any irrep.  An argument for the determination of the ranges of the angles in the Euler angle parameterization is given.  Finally the states within the octet are given and a description of the direct calculation of the Wigner Clebsch Gordon coefficients is given that uses the invariant volume element.

\section{The Ladder Operations}

The so-called ladder, or raising and lowering operators, take one state to another within an irrep.  Their representation may be in terms of matrices or differential operators.  The differential operators here have been constructed from linear combinations of the left invariant vector fields in \cite{me}.  This would enable one to analyze the states within a group representation.  Most of this analysis has been performed using the properties of the commutation relations which the differential operators can be shown to satisfy.  These analyses will not be repeated here since they are well explained in various texts (see for example \cite{gas} and \cite{gandm}).  What is important here is that the differential operators given can be shown to satisfy the commutation relations on the $D$ matrices of the next section and therefore represent the Lie algebra as claimed.  These follow.  

First the left differential operators, that is, those that are constructed from the left invariant vector fields of \cite{me}.  These also change the labels on the left of the brackets that I have used to represent the elements of the $D$ matrices.  In what follows 
\bea
\partial_1 &\equiv & \frac{\partial}{\partial \alpha}, \;\;\;\;\;\;\;\;\;\;\;\;\;\;\; 
\partial_2 \equiv \frac{\partial}{\partial \beta}, \;\;\;\;\;\;\;\;\;\;\;\;\;\;\;
\hfil \partial_3 \equiv \frac{\partial}{\partial \gamma}\nonumber \\
\partial_5 &\equiv& \frac{\partial}{\partial a},   \;\;\;\;\;\;\;\;\;\;\;\;\;\;\;
\partial_6 \equiv \frac{\partial}{\partial b}, \;\;\;\;\;\;\;\;\;\;\;\;\;\;\; 
\partial_7 \equiv  \frac{\partial}{\partial c},\;\;\;\;\;\;\;\;\;\;\;\;\;\;\;
\nonumber \\
\partial_4 &\equiv & \frac{\partial}{\partial \theta}, \;\;\;\;\;\;\;\;\;\;\;\;\;\;\;
\partial_8 \equiv \frac{\partial}{\partial \phi}.\;\;\;\;\;\;\;\;\;\;\;\;\;\;\;\nonumber 
\eea

\bea
T_+&=&\half(\Lambda_1 + i\Lambda_2) = \half e^{-2i\alpha}(i\cot2\beta \partial_1 -\partial_2 - \frac{i}{\sin2\beta}\partial_3)\\
T_-&=&\half(\Lambda_1 - i\Lambda_2) = \half e^{2i\alpha}(i\cot2\beta \partial_1 +\partial_2 
- \frac{i}{\sin2\beta}\partial_3)\\
V_+&=&\half(\Lambda_4 + i\Lambda_5)\nonumber\\ 
   &=&\frac{i}{2}e^{-i(\alpha + \gamma)}\frac{\sin \beta}{\sin2\beta}\cot \theta\partial_1
    + \half e^{-i(\alpha + \gamma)}\sin \beta \cot \theta \partial_2\nonumber\\
   &-&\frac{i}{2}e^{-i(\alpha + \gamma)}\cot 2\beta \sin \beta \cot \theta \partial_3
    + \frac{i}{2} e^{-i(\alpha + \gamma)}\frac{(2-\sin^2\theta)}{\sin 2\theta}\cos \beta \partial_3\nonumber\\
   &-& \half e^{-i(\alpha + \gamma)}\cos \beta \partial_4\nonumber\\
   &-&\frac{i}{2}e^{-i(\alpha + \gamma)}\frac{2\cos \beta}{\sin 2\theta}\partial_5
    - \frac{i}{2}e^{-i(\alpha - \gamma - 2a)}\frac{\cot 2b}{\sin \theta} \sin \beta \partial_5\nonumber\\
   &-& \half e^{-i(\alpha - \gamma - 2a)}\frac{\sin \beta}{\sin \theta} \partial_6
    + \frac{i}{2}e^{-i(\alpha - \gamma - 2a)}\frac{\sin \beta}{\sin \theta \sin 2b}\partial_7\nonumber\\
   &-& \frac{3}{4} e^{-i(\alpha + \gamma)}\tan \theta \cos \beta Y_8
\eea
\bea
V_-&=&\half(\Lambda_4 - i\Lambda_5)\nonumber\\
   &=&\frac{i}{2}e^{i(\alpha + \gamma)}\frac{\sin \beta}{\sin2\beta}\cot \theta\partial_1
    - \half e^{i(\alpha + \gamma)}\sin \beta \cot \theta \partial_2\nonumber\\
   &-&\frac{i}{2}e^{i(\alpha + \gamma)}\cot 2\beta \sin \beta \cot \theta \partial_3
    + \frac{i}{2}e^{i(\alpha + \gamma)}\frac{(2-\sin^2\theta)}{\sin 2\theta}\cos \beta \partial_3\nonumber\\
   &+& \half e^{i(\alpha + \gamma)}\cos \beta \partial_4\nonumber\\
   &-&\frac{i}{2}e^{i(\alpha + \gamma)}\frac{2\cos \beta}{\sin 2\theta}\partial_5
    - \frac{i}{2}e^{i(\alpha - \gamma - 2a)}\frac{\cot 2b}{\sin \theta} \sin \beta \partial_5\nonumber\\
   &+& \half e^{i(\alpha - \gamma - 2a)}\frac{\sin \beta}{\sin \theta} \partial_6
    + \frac{i}{2}e^{i(\alpha - \gamma - 2a)}\frac{\sin \beta}{\sin \theta \sin 2b}\partial_7\nonumber\\
   &-& \frac{3}{4} e^{i(\alpha + \gamma)}\tan \theta \cos \beta Y_8
\eea
\bea
U_+&=&\half(\Lambda_6 + i\Lambda_7)\nonumber\\
   &=&\frac{i}{2}e^{i(\alpha - \gamma)}\frac{\cos \beta}{\sin2\beta}\cot \theta\partial_1
    + \half e^{i(\alpha - \gamma)}\cos \beta \cot \theta \partial_2\nonumber\\
   &-&\frac{i}{2}e^{i(\alpha - \gamma)}\cot 2\beta \cos \beta \cot \theta \partial_3
    - \frac{i}{2}e^{i(\alpha - \gamma)}\frac{(2-\sin^2\theta)}{\sin 2\theta}\sin \beta \partial_3\nonumber\\
   &+& \half e^{i(\alpha - \gamma)}\sin \beta \partial_4\nonumber\\
   &+&\frac{i}{2}e^{i(\alpha - \gamma)}\frac{2\sin \beta}{\sin 2\theta}\partial_5
    - \frac{i}{2}e^{i(\alpha + \gamma + 2a)}\frac{\cot 2b}{\sin \theta} \cos \beta \partial_5\nonumber\\
   &-& \half e^{i(\alpha + \gamma + 2a)}\frac{\cos \beta}{\sin \theta} \partial_6
    + \frac{i}{2}e^{i(\alpha + \gamma + 2a)}\frac{\cos \beta}{\sin \theta \sin 2b}\partial_7\nonumber\\
   &+&\frac{3}{4} e^{i(\alpha - \gamma)}\tan \theta \sin \beta Y_8
\eea
\bea
U_-&=&\half(\Lambda_6 - i\Lambda_7)\nonumber\\
   &=&\frac{i}{2}e^{-i(\alpha - \gamma)}\frac{\cos \beta}{\sin2\beta}\cot \theta\partial_1
    - \half e^{-i(\alpha - \gamma)}\cos \beta \cot \theta \partial_2\nonumber\\
   &-&\frac{i}{2}e^{-i(\alpha - \gamma)}\cot 2\beta \cos \beta \cot \theta \partial_3
    - \frac{i}{2}e^{-i(\alpha - \gamma)}\frac{(2-\sin^2\theta)}{\sin 2\theta}\sin \beta \partial_3\nonumber\\
   &-& \half e^{-i(\alpha - \gamma)}\sin \beta \partial_4\nonumber\\
   &+&\frac{i}{2}e^{-i(\alpha - \gamma)}\frac{2\sin \beta}{\sin 2\theta}\partial_5
    - \frac{i}{2}e^{-i(\alpha + \gamma + 2a)}\frac{\cot 2b}{\sin \theta} \cos \beta \partial_5\nonumber\\
   &+& \half e^{-i(\alpha + \gamma + 2a)}\frac{\cos \beta}{\sin \theta} \partial_6
    + \frac{i}{2}e^{-i(\alpha + \gamma + 2a)}\frac{\cos \beta}{\sin \theta \sin 2b}\partial_7\nonumber\\
   &+& \frac{3}{4} e^{-i(\alpha - \gamma)}\tan \theta \sin \beta Y_8\\
T_3&=&\frac{i}{2}\partial_1\\
Y&=&i\partial_3 - i\partial_5 + i\frac{1}{\sqrt{3}}\partial_8
\eea
Where I have omitted a ``left'' designation.  The right differential operators have a superscript $r$.  These are given by the following equations.

\bea
T_-^r&=&\half(\Lambda_1^r + i\Lambda_2^r) = \half e^{2ic}(-i\cot 2b \partial_7 - \partial_6 + \frac{i}{\sin 2b}\partial_5)\\
T_+^r&=&\half(\Lambda_1^r - i\Lambda_2^r) = \half e^{-2ic}(-i\cot 2b \partial_7 + \partial_6 
+ \frac{i}{\sin 2b}\partial_5)\\
V_-^r&=&\half(\Lambda_4^r + i\Lambda_5^r)\nonumber\\ 
   &=&-\frac{i}{2}e^{i(c+a+3\eta)}\frac{\sin b}{\sin2b}\cot \theta\partial_7
    + \half e^{i(c+a+3\eta)}\sin b \cot \theta \partial_6\nonumber\\
   &+&\frac{i}{2}e^{i(c+a+3\eta)}\cot 2b \sin b \cot \theta \partial_5
    - \frac{i}{2}e^{i(c+a+3\eta)}\frac{(2-\sin^2\theta)}{\sin 2\theta}\cos b \partial_5\nonumber\\
   &-& \half e^{i(c+a+3\eta)}\cos b \partial_4\nonumber\\
   &+&\frac{i}{2}e^{i(c+a+3\eta)}\frac{2\cos b}{\sin 2\theta}\partial_3
    + \frac{i}{2}e^{i(c-a-2\gamma+3\eta)}\frac{\cot 2\beta}{\sin \theta} \sin b \partial_3\nonumber\\
   &-& \half e^{i(c-a-2\gamma+3\eta)}\frac{\sin b}{\sin \theta} \partial_2
    - \frac{i}{2}e^{i(c-a-2\gamma+3\eta)}\frac{\sin b}{\sin \theta \sin 2\beta}\partial_1\nonumber\\
   &+& \frac{3}{4} e^{i(c+a+3\eta)}\tan \theta \cos b Y_8^r
\eea
\bea
V_+^r&=&\half(\Lambda_4^r - i\Lambda_5^r)\nonumber\\ 
   &=&-\frac{i}{2}e^{-i(c+a+3\eta)}\frac{\sin b}{\sin2b}\cot \theta\partial_7
    - \half e^{-i(c+a+3\eta)}\sin b \cot \theta \partial_6\nonumber\\
   &+&\frac{i}{2}e^{-i(c+a+3\eta)}\cot 2b \sin b \cot \theta \partial_5
    - \frac{i}{2}e^{-i(c+a+3\eta)}\frac{(2-\sin^2\theta)}{\sin 2\theta}\cos b \partial_5\nonumber\\
   &+& \half e^{-i(c+a+3\eta)}\cos b \partial_4\nonumber\\
   &+&\frac{i}{2}e^{-i(c+a+3\eta)}\frac{2\cos b}{\sin 2\theta}\partial_3
    + \frac{i}{2}e^{-i(c-a-2\gamma+3\eta)}\frac{\cot 2\beta}{\sin \theta} \sin b \partial_3\nonumber\\
   &+& \half e^{-i(c-a-2\gamma+3\eta)}\frac{\sin b}{\sin \theta} \partial_2
    - \frac{i}{2}e^{-i(c-a-2\gamma+3\eta)}\frac{\sin b}{\sin \theta \sin 2\beta}\partial_1\nonumber\\
   &+& \frac{3}{4} e^{-i(c+a+3\eta)}\tan \theta \cos b Y_8^r
\eea
\bea
U_-^r&=&\half(\Lambda_6^r + i\Lambda_7^r)\nonumber\\
   &=&\frac{i}{2}e^{-i(c-a-3\eta)}\frac{\cos b}{\sin 2b}\cot \theta\partial_7
    - \half e^{-i(c-a-3\eta)}\cos b \cot \theta \partial_6\nonumber\\
   &-&\frac{i}{2}e^{-i(c-a-3\eta)}\cot 2b \cos b \cot \theta \partial_5
    - \frac{i}{2}e^{-i(c-a-3\eta)}\frac{(2-\sin^2\theta)}{\sin 2\theta}\sin b \partial_5\nonumber\\
   &-& \half e^{-i(c-a-3\eta)}\sin b \partial_4\nonumber\\
   &+&\frac{i}{2}e^{-i(c-a-3\eta)}\frac{2\sin b}{\sin 2\theta}\partial_3
    - \frac{i}{2}e^{-i(c+a+2\gamma - 3\eta)}\frac{\cot 2\beta}{\sin \theta} \cos b \partial_3\nonumber\\
   &+& \half e^{-i(c+a+2\gamma - 3\eta)}\frac{\cos b}{\sin \theta} \partial_2
    + \frac{i}{2}e^{-i(c+a+2\gamma - 3\eta)}\frac{\cos b}{\sin \theta \sin 2\beta}\partial_1\nonumber\\
   &+& \frac{3}{4} e^{-i(c-a-3\eta)}\tan \theta \sin b Y_8^r
\eea
\bea
U_+^r&=&\half(\Lambda_6^r - i\Lambda_7^r)\nonumber\\
   &=&\frac{i}{2}e^{i(c-a-3\eta)}\frac{\cos b}{\sin 2b}\cot \theta\partial_7
    + \half e^{i(c-a-3\eta)}\cos b \cot \theta \partial_6\nonumber\\
   &-&\frac{i}{2}e^{i(c-a-3\eta)}\cot 2b \cos b \cot \theta \partial_5
    - \frac{i}{2}e^{i(c-a-3\eta)}\frac{(2-\sin^2\theta)}{\sin 2\theta}\sin b \partial_5\nonumber\\
   &+& \half e^{i(c-a-3\eta)}\sin b \partial_4\nonumber\\
   &+&\frac{i}{2}e^{i(c-a-3\eta)}\frac{2\sin b}{\sin 2\theta}\partial_3
    - \frac{i}{2}e^{i(c+a+2\gamma - 3\eta)}\frac{\cot 2\beta}{\sin \theta} \cos b \partial_3\nonumber\\
   &-& \half e^{i(c+a+2\gamma - 3\eta)}\frac{\cos b}{\sin \theta} \partial_2
    + \frac{i}{2}e^{i(c+a+2\gamma - 3\eta)}\frac{\cos b}{\sin \theta \sin 2\beta}\partial_1\nonumber\\
   &+& \frac{3}{4} e^{i(c-a-3\eta)} \tan \theta \sin b Y_8^r\\
T_3^r&=&\frac{i}{2}\partial_7\\
Y^r&=&\frac{1}{\sqrt{3}}i\partial_8
\eea

where $\eta \equiv \phi/\sqrt{3}$.  These operations are given explicitly by example below.  One may take note that the right ``raising'' operations are given by the subtraction of two elements of the corresponding $\Lambda$s.  This is due to the commutation relations that are obeyed by the right operators.  They satisfy \cite{lb}
$$
[\Lambda_i,\Lambda_j] = -2i \epsilon_{ijk} \Lambda_k,
$$
whereas the left operators satisfy
$$
[\Lambda_i,\Lambda_j] = 2i \epsilon_{ijk} \Lambda_k.
$$

\section{The Fundamental Representations}

The fundamental representations can be obtained from the parameterization in \cite{me} by direct exponentiation and multiplication or by the general expression for the maximum weight state in the next section.  The maximum weight state can be derived from the differential operators above as will be explained in the next section.  Once obtained, the maximum weight state can be used to find every other state within a given irrep by operation with the raising and lowering operators.  Here the fundamental representations are exhibited explicitly and one may check through straight forward calculation that they are related through the general operations defined above.

First the  ${\bf 3}$ representation.
\beq
D(\alpha,\beta,\gamma,\theta,a,b,c,\phi) = e^{(-i\la_3 \alpha)} e^{(i\la_2 \beta)}
 e^{(-i\la_3 \gamma)} e^{(i\la_5 \theta)} e^{(-i\la_3 a)} e^{(i\la_2 b)} 
e^{(-i\la_3 c)} e^{(-i\la_8 \phi)},
\eeq
This matrix actually corresponds to the complex conjugate of the matrix $D$ in \cite{me} as is common.  The particular signs of the exponents correspond to a choice of phase that is a generalization of the Condon and Shortley phase convention (see \cite{BandBIII}).  This makes the root operators positive or zero.  Matrix elements can be labeled by their eigenvalues as below.  Where the following definition is used:
$$
<t_3^\prime,y^\prime||t_3,y> \equiv D^{(1,0)}_{t_3,y;t_3^{\prime},y^{\prime}}
$$
\beq
D^{(1,0)}_{t_3,y;t_3^{\prime},y^{\prime}} =  \nonumber \\
\left( \begin{array}{crcl}
<\half,\frac{1}{3}||\half,\frac{1}{3}> 
& <\half,\frac{1}{3}||-\half,\frac{1}{3}> 
& <\half,\frac{1}{3}||0,-\frac{2}{3}> \\
<-\half,\frac{1}{3}||\half,\frac{1}{3}> 
& <-\half,\frac{1}{3}||-\half,\frac{1}{3}> 
& <-\half,\frac{1}{3}||0,-\frac{2}{3}> \\
<0,-\frac{2}{3}||\half,\frac{1}{3}> 
& <0,-\frac{2}{3}||-\half,\frac{1}{3}> 
& <0,-\frac{2}{3}||0,-\frac{2}{3}>   
\end{array} \right)
\eeq
These matrix elements correspond to functions:
\beq
<\half,\frac{1}{3}||\half,\frac{1}{3}>=
e^{-i\alpha}e^{-ic}e^{-i\eta}(e^{-i\gamma}e^{-ia}\cos \beta \cos b \cos \theta 
-e^{i\gamma}e^{ia}\sin \beta \sin b) 
\eeq
\beq
<\half,\frac{1}{3}||-\half,\frac{1}{3}> = e^{-i\alpha}e^{ic}e^{-i\eta}
(e^{-i\gamma}e^{-ia}\cos \beta \sin b \cos \theta 
+e^{i\gamma}e^{ia}\sin \beta \cos b) 
\eeq
\beq
<\half,\frac{1}{3}||0,-\frac{2}{3}>
= e^{-i\alpha}e^{-i\gamma}e^{2i\eta}\cos \beta \sin \theta 
\eeq
\beq
<-\half,\frac{1}{3}||\half,\frac{1}{3}>
= -e^{i\alpha}e^{-ic}e^{-i\eta}(e^{-i\gamma}e^{-ia}\sin \beta \cos b \cos \theta 
+e^{i\gamma}e^{ia}\cos \beta \sin b) 
\eeq
\beq
<-\half,\frac{1}{3}||-\half,\frac{1}{3}>
= -e^{i\alpha}e^{ic}e^{-i\eta}(e^{-i\gamma}e^{-ia}\sin \beta \sin b \cos \theta -e^{i\gamma}e^{ia}\cos \beta \cos b) 
\eeq
\beq
<-\half,\frac{1}{3}||0,-\frac{2}{3}>
= -e^{i\alpha}e^{-i\gamma}e^{2i\eta}\sin \beta \sin \theta 
\eeq
\beq
<0,-\frac{2}{3}||\half,\frac{1}{3}>
= -e^{-ia}e^{-ic}e^{-i\eta}\sin \theta \cos b 
\eeq
\beq
<0,-\frac{2}{3}||-\half,\frac{1}{3}>
= -e^{-ia}e^{ic}e^{-i\eta} \sin b \sin \theta
\eeq
\beq
<0,-\frac{2}{3}||0,-\frac{2}{3}>
= e^{2i\eta}\cos \theta   
\eeq
\hfil\break

This is actually formed from $D^*$ and the ${\bf 3}^*$ representation if formed by the following replacements:  \{$\la_1$, $-\la_2$, $-\la_3$, $\la_4$, $-\la_5$, $\la_6$, $-\la_7$, $-\la_8$\} for the corresponding matrices in the ${\bf 3}$ representation.  The two fundamental representations are inequivalent so there exists no inner automorphism between them.  This is the outer automorphism that preserves the ladder operations and preserves the previous phase convention.  The ${\bf 3}^*$ representation is then found to be as follows.

\beq
D(\alpha,\beta,\gamma,\theta,a,b,c,\phi) = e^{(i\la_3 \alpha)} e^{(-i\la_2 \beta)}
 e^{(i\la_3 \gamma)} e^{(-i\la_5 \theta)} e^{(i\la_3 a)} e^{(-i\la_2 b)} 
e^{(i\la_3 c)} e^{(i\la_8 \phi)},
\eeq
And its matrix elements can be labeled by its corresponding eigenvalues as follows.
\beq
D^{(0,1)}_{t_3,y;t_3^{\prime},y^{\prime}} = \left( \begin{array}{crcl}
<-\half,-\frac{1}{3}||-\half,-\frac{1}{3}> 
& <-\half,-\frac{1}{3}||\half,-\frac{1}{3}> 
& <-\half,-\frac{1}{3}||0,\frac{2}{3}> \\
<\half,-\frac{1}{3}||-\half,-\frac{1}{3}> 
& <\half,-\frac{1}{3}||\half,-\frac{1}{3}> 
& <\half,-\frac{1}{3}||0,\frac{2}{3}> \\
<0,\frac{2}{3}||-\half,-\frac{1}{3}> 
& <0,\frac{2}{3}||\half,-\frac{1}{3}> 
& <0,\frac{2}{3}||0,\frac{2}{3}>   \end{array} \right)
\eeq
\beq
<-\half,-\frac{1}{3}||-\half,-\frac{1}{3}>
= e^{i\alpha}e^{ic}e^{i\eta}(e^{i\gamma}e^{ia}\cos \beta \cos b \cos \theta
-e^{-i\gamma}e^{-ia}\sin \beta \sin b)
\eeq
\beq
<-\half,-\frac{1}{3}||\half,-\frac{1}{3}>
= -e^{i\alpha}e^{-ic}e^{i\eta}(e^{i\gamma}e^{ia}\cos \beta \sin b \cos \theta
+e^{-i\gamma}e^{-ia}\sin \beta \cos b)
\eeq
\beq
<-\half,-\frac{1}{3}||0,\frac{2}{3}>
=-e^{i\alpha}e^{i\gamma}e^{-2i\eta}\cos \beta \sin \theta
\eeq
\beq
<\half,-\frac{1}{3}||-\half,-\frac{1}{3}>
= e^{-i\alpha}e^{ic}e^{i\eta}(e^{i\gamma}e^{ia}\sin \beta \cos b \cos \theta
+e^{-i\gamma}e^{-ia}\cos \beta \sin b)
\eeq
\beq
<\half,-\frac{1}{3}||\half,-\frac{1}{3}>
= -e^{-i\alpha}e^{-ic}e^{i\eta}(e^{i\gamma}e^{ia}\sin \beta \sin b \cos \theta
-e^{-i\gamma}e^{-ia}\cos \beta \cos b)
\eeq
\beq
<\half,-\frac{1}{3}||0,\frac{2}{3}>
= -e^{-i\alpha}e^{i\gamma}e^{-2i\eta}\sin \beta \sin \theta
\eeq
\beq
<0,\frac{2}{3}||-\half,-\frac{1}{3}>
= e^{ia}e^{ic}e^{i\eta}\sin \theta \cos b
\eeq
\beq
<0,\frac{2}{3}||\half,-\frac{1}{3}>
= -e^{ia}e^{-ic}e^{i\eta}\sin \theta \sin b
\eeq
\beq
<0,\frac{2}{3}||0,\frac{2}{3}>
= e^{-2i\eta} \cos \theta
\eeq

\section{Irreducible Representations}

For each irrep there exists a maximum weight, $D^{(p,q)}_m$, state that can be defined by the following equations.
$$
V_+D^{(p,q)}_m = 0, \;\;\;\;\;\;\;\;\; V_+^rD^{(p,q)}_m = 0,
$$
$$
U_+D^{(p,q)}_m = 0, \;\;\;\;\;\;\;\;\; U_+^rD^{(p,q)}_m = 0,
$$
$$
T_+D^{(p,q)}_m = 0, \;\;\;\;\;\;\;\;\; T_+^rD^{(p,q)}_m = 0. 
$$
When one solves these equations and satisfies the conditions for the first two 
or three reps, one finds that in this parameterization
\bea
D^{(p,q)}_m & = & e^{-i(2q+p)\eta}e^{-ip\alpha}e^{-ipc}\sum_{n=0}^p (-1)^{n+1}
                        \left( \begin{array}{c} p\\n 
\end{array}          \right)\\
            &   & \times (e^{-i\gamma}e^{-ia} \cos \beta \cos b \cos \theta)^n
(e^{i\gamma}e^{ia} \sin \beta \sin b)^{p-n}\cos^q\theta.
\eea

{\it Note 1:  This is not the maximum state defined in \cite{gas} and 
\cite{gandm}}.  

The maximum state could also be labelled with $t_{3m}$ and $y_m$, which stand for the value of $t_3$ and $y$ for this maximum state.  In terms of $p$ and $q$ these are
$$
y_m = \frac{2q+p}{3},\;\;\;\;\;\;\;\;\;\;\; t_{3m} = \frac{p}{2}.
$$

\section{The Octet}

The octet is the smallest nontrivial example within which there exists two different states with the same $t_3$ and $y$.  These will have different $T$ of total isospin since they belong to different isospin representations.  Thus it may be used as an example of how the Clebsch-Gordon coefficients may be found using the explicit $D$ matrices.  

The octet is an irrep with eight states (hence the name).  It can be obtained from a product of $D^{(1,0)}$ and $D^{(0,1)}$ from which a scalar $D^{(0,0)}$ is removed.  Thus it is denoted $D^{(1,1)}$.  For it, the maximum weight state is given by the equation in the last section by substitution of the explicit $p$ and $q$.  
$$
D^{(1,1)}_m = e^{-i\alpha} e^{-ic} e^{-3i\eta} \cos(\theta)[e^{-i\gamma} e^{-ia} \cos(\beta) \cos(b) \cos(\theta) - e^{i\gamma} e^{ia} \sin(\beta) \sin(b)].
$$
For calculational purposes it is more convenient to notice that this may be written as
$$
D^{(1,1)}_m =<\half,\frac{1}{3}||\half,\frac{1}{3}><0,\frac{2}{3}||0,\frac{1}{3}>.
$$
From this state, operation by $V_-$ will give one of the two different center states, each having $(t_3,y)$ given by $(0,0)$.  The first is given by
$$
V_-D^{(1,1)}_m = <-\half,\frac{1}{3}||\half,\frac{1}{3}><\half,-\frac{1}{3}||0,\frac{2}{3}>,
$$
and the second by
\bea
T_-U_-D^{(1,1)}_m &=& <-\half,\frac{1}{3}||\half,\frac{1}{3}><\half,-\frac{1}{3}||0,\frac{2}{3}> \nonumber \\
&+& <\half,\frac{1}{3}||\half,\frac{1}{3}><-\half,-\frac{1}{3}||0,\frac{2}{3}> \nonumber
\eea
The other states are as follows, listed counter clockwise around the hexagon starting from the one after the maximum weight state.
$$
U_-D^{(1,1)}_m =  <\half,\frac{1}{3}||\half,\frac{1}{3}><\half,-\frac{1}{3}||0,\frac{2}{3}>.
$$
$$
V_-U_-D^{(1,1)}_m =  <0,-\frac{2}{3}||\half,\frac{1}{3}><\half,-\frac{1}{3}||0,\frac{2}{3}>.
$$
$$
T_-V_-U_-D^{(1,1)}_m =  <0,-\frac{2}{3}||\half,\frac{1}{3}><-\half,-\frac{1}{3}||0,\frac{2}{3}>. 
$$
$$
U_+T_-V_-U_-D^{(1,1)}_m =  <-\half,\frac{1}{3}||\half,\frac{1}{3}><-\half,-\frac{1}{3}||0,\frac{2}{3}>.
$$
$$
V_+U_+T_-V_-U_-D^{(1,1)}_m =  <-\half,\frac{1}{3}||\half,\frac{1}{3}><0,\frac{2}{3}||0,\frac{2}{3}>.
$$
The two of concern here are the two center states.  From these, the linear combinations that give states that are members of $SU(2)$ isospin states will be used.
This is easy to do.  Simply take the an arbitrary linear combination of the two and demand that on this state $T_+$ and $T_-$ on it give zero.  This linear combination is then a member of an isospin singlet.  The other linear combination gives the center state in an isospin triplet.  These linear combinations are found to be
$$
D^{(1,1)}_{(2,0,0;2,0,0)} = <-\half,\frac{1}{3}||\half,\frac{1}{3}><\half,-\frac{1}{3}||0,\frac{2}{3}>,
$$
which is the member of the isospin triplet, and
\bea
D^{(1,1)}_{(0,0,0;0,0,0)} &=& <-\half,\frac{1}{3}||\half,\frac{1}{3}><\half,-\frac{1}{3}||0,\frac{2}{3}> \nonumber \\
&-& <\half,\frac{1}{3}||\half,\frac{1}{3}><-\half,-\frac{1}{3}||0,\frac{2}{3}>,
\eea
which is an isospin singlet.

\hfil\break

  The $D$ matrices are labelled properly in the following form (the $t$ label was not neccesary in the fundamental representations nor would it be on any triangular representation, $D^{(p,0)}$ or $D^{(0,q)}$.)
$$
D^{(p,q)}_{(t,t_3,y;t^{\prime},t_3^{\prime},y^{\prime})}
$$
 Thus the Clebsch-Gordon coefficients have been determined.  This can be used as a general method for calculating them.  One can simply demand that the states form complete horizontal isospin irreps in the $t_3-y$ plane.  These are not $SU(3)$ WCG (Wigner-Clebsch-Gordon) coefficients, but rather the coefficients of the linear combinations of $SU(2)$ irreps within $SU(3)$.  The method of calculating the $SU(3)$ WCG coefficients is now straight forward and will be discussed next.

\section{WCG Coefficients for $SU(3)$}

The WCG coefficients may now be calculated with the orthogonality relations between different states using the following group invariant volume element.  This may be found by using the (wedge) product of the left(or right) invariant one forms calculated in \cite{me}.  The result is the following

$$
dV = \sin 2\beta \sin 2b \sin 2\theta \sin^2 \theta\; d\alpha\; d\beta\; d\gamma\;
d\theta\; da\; db\; dc\; d\phi,
$$
where the ranges of integration are
$$
0 \leq \alpha,\gamma,a,c < \pi, 
$$
$$
 0 \leq \beta,b,\theta \leq \frac{\pi}{2} \;\;\;\;\mbox{and}\;\;\;\; 0 \leq \phi < \sqrt{3}\pi.
$$
These are not trivial to determine \cite{Marinov} since their determination is equivalent to determining the invariant volume of the group.  With the $D$ matrices given for the fundamental representations, one may infer these minimum values for the ranges of the angles by enforcing the orthogonality relations that these representation functions must satisfy.  These orthogonality relations are given by

\bea
\int D^{{(p_1,q_1)}*}_{t_1,(t_3)_1,y_1;t^{\prime}_1,(t^{\prime}_3)_1,y^{\prime}_1}  D^{(p_2,q_2)}_{t_2,(t_3)_2,y_2;t^{\prime}_2,(t^{\prime}_3)_2,y^{\prime}_2}dV
&=&\nonumber\\ 
\frac{V_0}{d}\delta_{p_1,p_2}\delta_{q_1,q_2}\delta_{t_1,t_2}\delta_{(t_3)_1,(t_3)_2}
\delta_{y_1,y_2}\delta_{t^{\prime}_1,t^{\prime}_2}
\delta_{(t^{\prime}_3)_1,(t^{\prime}_3)_2}\delta_{y^{\prime}_1,y^{\prime}_2},
\eea
where $V_0$ is the invariant volume of the group and $d$ is the dimension of the representation, $d = \frac{1}{2}(p+1)(q+1)(p+q+2)$.  Thus knowing that the integral of product of an element of a $D$ matrix with its complex conjugate is a constant that depends only on the dimensionality of the representation, and that the integral of its product with anything else is zero, provides equations that may be solved to find the ranges of the angles.  The result for $V_0$ agrees with what Marinov found within a factor of 3 \cite{Marinov}.  This may be explained by considering the structure of the group manifold.  In reference \cite{lb} the group invariant volume element for $SU(2)$ is derived.  The nomalization factor $\pi^2$ can be viewed as arising from the angles $\alpha$, and $\beta$ in the ordinary Euler angle parameterization of $SU(2)$;
$$
U = e^{i\alpha J_3} e^{i\beta J_2} e^{i\gamma J_3}.
$$
The factor of $2$ comes from the covering of the northern and southern poles, or hemispheres.  In the case of $SU(3)$ one could consider the possibility of $3$ ``poles''.  Some evidence for this is exhibited by the othogonality relations for the density matrix of the 3 state systems considered in \cite{m2}.

The orthogonality relation for the $SU(3)$ representation matrices, with the constants determined, gives us a vital tool for the determination of the WCG coefficients of $SU(3)$.  One may simply take a direct product of any number of the fundamental representations and use the orthogonality relation to determine the representations contained in that direct product and how many of them there are.  The linear combinations of the states in a given representation can then be determined with the coefficients being WCG coefficients either by direct operation with the raising and lowering operators that were given earlier, or by ensuring othogonality with the appropriate integration and the states generated by the highest weight state given in section 4.  These are equivalent and hence nothing new.  What {\ is} new, is the orthogonality relation with appropriate constants.  This eliminates the problem faced by deSwart by solving his $\Gamma$ problem \cite{deSwart}.

\section{$SU(3)$ and $SO(8)$}

The generic element of the adjoint representation, since it is real and unitary, is orthogonal.  Since it also has determinant $1$, it is an element of $SO(8)$.  It is however, a function of only eight angles and so is a special element.  If we call this matrix $R_{ij}$, then it will satisfy the equation
$$
\Lambda_i^r = R_{ij}\Lambda_j,
$$
or,
$$
U\lambda_iU^{\dagger} =  R_{ij}\lambda_j,
$$
Therefore we have a mapping from the left invariant vector fields to the right invariant vector fields.  

This mapping is exhibited explicitly here.

\bea
R_{11} &=& \cos 2\alpha \cos2\beta \cos \theta [ \cos(2a+2\gamma) \cos 2b\cos 2c-\sin(2a+2\gamma)\sin 2c]  \nonumber \\
       &-& \sin 2\alpha \cos \theta [ \sin(2a+2\gamma) \cos 2b \cos 2c + \cos(2a+2\gamma) \sin 2c] \nonumber \\
       &-& \cos 2\alpha\sin 2\beta(1-\half\sin^2 \theta)\sin 2b \cos 2c 
\eea
\bea
R_{12} &=& \sin 2\alpha \cos2\beta \cos \theta [ \cos(2a + 2\gamma) \cos 2b\cos 2c - \sin(2a + 2\gamma)\sin 2c]  \nonumber \\
       &+& \cos 2\alpha \cos \theta [ \sin (2a + 2\gamma) \cos 2b \cos 2c + \cos(2a + 2\gamma) \sin 2c] \nonumber \\
       &-& \sin 2\alpha\sin 2\beta (1-\half \sin^2 \theta)\sin 2b \sin 2c 
\eea
\bea
R_{13} &=& \sin 2\beta \cos (2a +2\gamma) \cos 2b \cos 2c \cos \theta - \sin 2\beta \sin(2a + 2\gamma) \sin 2c \cos \theta  \nonumber \\
       &+& \cos 2\beta (1-\half \sin^2 \theta) \sin 2b \cos 2c
\eea
\bea
R_{14} &=&-\half \cos (\alpha+\gamma) \cos \beta \sin 2\theta \sin 2b \cos 2c 
\nonumber \\
       &-& \cos(\alpha - \gamma -2a)\sin \beta \cos 2b \cos 2c \sin \theta
  \nonumber \\
       &+& \sin(\alpha + \gamma +2a)\sin \beta \sin 2c \sin \theta
\eea
\bea
R_{15} &=& \half \sin(\alpha+\gamma) \cos \beta \sin 2\theta \sin 2b \cos 2c 
\nonumber \\
       &+& \sin(\alpha - \gamma -2a)\sin \beta \cos 2b \cos 2c \sin \theta
  \nonumber \\
       &+& \cos(\alpha + \gamma +2a)\sin \beta \sin 2c \sin \theta
\eea
\bea
R_{16} &=& \half \cos(\alpha-\gamma) \sin \beta \sin 2\theta \sin 2b \cos 2c 
\nonumber \\
       &-& \cos(\alpha - \gamma -2a)\cos \beta \cos 2b \cos 2c \sin \theta
  \nonumber \\
       &+& \sin(\alpha + \gamma +2a)\cos \beta \sin 2c \sin \theta
\eea
\bea
R_{17} &=& \half \sin(\alpha-\gamma) \sin \beta \sin 2\theta \sin 2b \cos 2c  
\nonumber \\
       &-& \sin(\alpha-\gamma-2a)\cos \beta \cos 2b \cos 2c \sin \theta
  \nonumber \\
       &-& \cos(\alpha+\gamma +2a)\cos \beta \sin 2c \sin \theta
\eea
\beq
R_{18} = -\frac{\sqrt{3}}{2}\sin^2 \theta \sin 2b \cos 2c
\eeq

\bea
R_{21} &=& \cos 2\alpha \cos2\beta \cos \theta [\sin(2a+2\gamma) \cos 2c + 
\cos(2a+2\gamma) \cos 2b \sin 2c]  \nonumber \\
       &-& \sin 2\alpha \cos \theta [\sin(2a+2\gamma) \cos 2b \sin 2c - \cos(2a+2\gamma) \cos 2c] \nonumber \\
       &-& \cos 2\alpha \sin 2\beta (1-\half\sin^2 \theta) \sin 2b \sin 2c 
\eea
\bea
R_{22} &=& -\sin 2\alpha \cos 2\beta \cos \theta [\sin(2a+2\gamma) \cos 2c + \cos(2a+2\gamma)\cos 2b \sin 2c]  \nonumber \\
       &-& \cos 2\alpha \cos \theta [\sin(2a+2\gamma) \cos 2b \sin 2c - \cos(2a+2\gamma) \cos 2c] \nonumber \\
       &+& \sin 2\alpha \sin 2\beta (1-\half\sin^2 \theta) \sin 2b \sin 2c 
\eea
\bea
R_{23} &=& \sin 2\beta \cos \theta[\cos(2a+2\gamma) \cos 2b \sin 2c + \sin(2a+2\gamma) \cos 2c]  \nonumber \\
       &+& \cos 2\beta (1-\half\sin^2 \theta) \sin 2b \sin 2c
\eea
\bea
R_{24} &=&-\half \cos(\alpha+\gamma) \cos \beta \sin 2\theta \sin 2b \sin 2c 
\nonumber \\
       &+& \sin(\alpha-\gamma-2a)\sin \beta \sin \theta \cos 2c
  \nonumber \\
       &-& \cos(\alpha-\gamma-2a)\sin \beta \sin \theta \cos 2b \sin 2c
\eea
\bea
R_{25} &=& \half \sin(\alpha+\gamma) \cos \beta \sin 2\theta \sin 2b \sin 2c 
\nonumber \\
       &+& \cos(\alpha-\gamma-2a) \sin \beta \sin \theta \cos 2c
  \nonumber \\
       &+& \sin(\alpha-\gamma-2a) \sin \beta \sin \theta \cos 2b \sin 2c
\eea
\bea
R_{26} &=& \half \cos (\alpha-\gamma) \sin \beta \sin 2\theta \sin 2b \sin 2c 
\nonumber \\
       &-& \sin(\alpha+\gamma+2a) \cos \beta \sin \theta \cos 2c
  \nonumber \\
       &-& \cos(\alpha+\gamma+2a)\cos \beta \sin \theta \cos 2b \sin 2c
\eea
\bea
R_{27} &=& \half \sin(\alpha-\gamma) \sin \beta \sin 2\theta \sin 2b \sin 2c 
\nonumber \\
       &+& \cos(\alpha+\gamma+2a) \cos \beta \sin \theta \cos 2c
  \nonumber \\
       &-& \sin(\alpha+\gamma+2a) \cos \beta \sin \theta \cos 2b \sin 2c
\eea
\beq
R_{28} = -\frac{\sqrt{3}}{2} \sin^2 \theta \sin 2b \sin 2c
\eeq

\bea
R_{31} &=& -\cos 2\alpha \cos 2\beta \cos \theta \sin 2b \cos(2a+2\gamma) 
       \nonumber \\
       &+& \sin 2\alpha \cos \theta \sin 2b \sin(2a+2\gamma)\nonumber \\
       &-& \cos 2\alpha \sin 2\beta (1-\half\sin^2\theta) \cos 2b 
\eea
\bea
R_{32} &=& \sin 2\alpha \cos2\beta \cos \theta \sin 2b \cos(2a+2\gamma) 
\nonumber \\
       &+& \cos 2\alpha \cos \theta \sin 2b \sin(2a+2\gamma)\nonumber \\
       &+& \sin 2\alpha \sin 2\beta (1-\half\sin^2\theta) \cos 2b 
\eea
\beq
R_{33} = -\sin 2\beta \cos \theta \sin 2b \cos(2a+2\gamma) + \cos 2\beta (1-\half\sin^2\theta)\cos 2b
\eeq
\beq
R_{34} = -\half \cos(\alpha +\gamma)\cos \beta \sin 2\theta \cos 2b 
+ \cos(\alpha-\gamma-2a) \sin \beta \sin \theta \sin 2b
\eeq
\beq
R_{35} = \half \sin(\alpha+\gamma)\cos \beta \sin 2\theta \cos 2b 
- \sin(\alpha-\gamma-2a) \sin \beta \sin \theta \sin 2b
\eeq
\beq
R_{36} = \half \cos(\alpha-\gamma)\sin \beta \sin 2\theta \cos 2b 
+ \cos(\alpha+\gamma+2a) \cos \beta \sin \theta \sin 2b
\eeq

\beq
R_{37} = \half \sin(\alpha-\gamma)\sin \beta \sin 2\theta \cos 2b + \sin(\alpha+\gamma+2a) \cos \beta \sin \theta \sin 2b
\eeq
\beq
R_{38} = -\frac{\sqrt{3}}{2}\sin^2 \theta \cos 2b
\eeq

\bea
R_{41} &=& -\cos 2\alpha \cos 2\beta \sin \theta \sin b \cos(a-c-2\gamma-3\eta)
\nonumber \\
	&-& \cos 2\alpha \sin 2\beta \sin 2\theta \cos(a+c+3\eta) \cos b
\nonumber \\
	&+& \sin 2\alpha \sin \theta \sin b \sin(a-c-2\gamma-3\eta)
\eea
\bea
R_{42} &=& \sin 2\alpha \cos 2\beta \sin \theta \sin b \cos(a-c-2\gamma-3\eta)
\nonumber \\
	&+& \sin 2\alpha \sin 2\beta \sin 2\theta \cos(a+c+3\eta) \cos b
\nonumber \\
	&-& \cos 2\alpha \sin \theta \sin b \sin(a-c-2\gamma-3\eta)
\eea
\bea
R_{43} &=& \sin 2\beta \sin \theta \sin b \cos(a-c-2\gamma-3\eta)
\nonumber \\
	&+& \cos 2\beta \sin 2\theta \cos(a+c+3\eta) \cos b
\eea
\bea
R_{44} &=& \cos(\alpha+\gamma) \cos \beta \cos 2\theta \cos(a+c+3\eta) \cos b
\nonumber \\
	&-& \sin(\alpha+\gamma) \cos \beta \sin(a+c+3\eta) \cos b
\nonumber \\
	&-& \sin \beta \sin \theta \sin b \cos(a+\gamma-\alpha-c-3\eta)
\eea
\bea
R_{45} &=& -\sin(\alpha+\gamma) \cos \beta \cos 2\theta \cos(a+c+3\eta) \cos b
\nonumber \\
	&-& \cos(\alpha+\gamma) \cos \beta \sin(a+c+3\eta) \cos b
\nonumber \\
	&-& \sin \beta \sin \theta \sin b \sin(a+\gamma-\alpha-c-3\eta)
\eea
\bea
R_{46} &=& -\cos(\alpha-\gamma) \sin \beta \cos 2\theta \cos(a+c+3\eta) \cos b
\nonumber \\
	&-& \sin(\alpha-\gamma) \sin \beta \sin(a+c+3\eta) \cos b
\nonumber \\
	&-& \cos \beta \cos \theta \sin b \cos(a+\gamma+\alpha-c-3\eta)
\eea
\bea
R_{47} &=& -\sin(\alpha-\gamma) \sin \beta \cos 2\theta \cos(a+c+3\eta) \cos b
\nonumber \\
	&+& \cos(\alpha-\gamma) \sin \beta \sin(a+c+3\eta) \cos b
\nonumber \\
	&-& \cos \beta \cos \theta \sin b \sin(a+\gamma+\alpha-c-3\eta)
\eea
\beq
R_{48} = \sqrt{3} \sin 2\theta \cos(a+c+3\eta) \cos b
\eeq

\bea
R_{51} &=& \cos 2\alpha \cos 2\beta \sin \theta \sin b \sin(a-c-2\gamma-3\eta)
\nonumber \\
       &-& \cos 2\alpha \sin 2\beta \sin 2\theta \sin(a+c+3\eta) \cos b
\nonumber \\
	&-& \sin 2\alpha \sin \theta \sin b \cos(a-c-2\gamma-3\eta)
\eea
\bea
R_{52} &=& -\sin 2\alpha \cos 2\beta \sin \theta \sin b \sin(a-c-2\gamma-3\eta)
\nonumber \\
	&+& \sin 2\alpha \sin 2\beta \sin 2\theta \sin(a+c+3\eta) \cos b
\nonumber \\
	&-& \cos 2\alpha \sin \theta \sin b \cos(a-c-2\gamma-3\eta)
\eea
\bea
R_{53} &=& \sin 2\beta \sin \theta \sin b \sin(a-c-2\gamma-3\eta)
\nonumber \\
       &+& \cos 2\beta \sin 2\theta \cos b \sin(a+c+3\eta)
\eea
\bea
R_{54} &=& \cos(\alpha+\gamma) \cos \beta \cos 2\theta \sin(a+c+3\eta) \cos b
\nonumber \\
       &+& \sin(\alpha+\gamma) \cos \beta \cos(a+c+3\eta) \cos b
\nonumber \\
       &+& \sin \beta \cos \theta \sin b \sin(a+\gamma-\alpha-c-3\eta)
\eea
\bea
R_{55} &=& -\sin(\alpha+\gamma) \cos \beta \cos 2\theta \sin(a+c+3\eta) \cos b
\nonumber \\
       &+& \cos(\alpha+\gamma) \cos \beta \cos(a+c+3\eta) \cos b
\nonumber \\
       &-& \sin \beta \cos \theta \sin b \cos(a+\gamma-\alpha-c-3\eta)
\eea
\bea
R_{56} &=& -\cos(\alpha-\gamma) \sin \beta \cos 2\theta \sin(a+c+3\eta) \cos b
\nonumber \\
       &+& \sin(\alpha-\gamma) \sin \beta \cos(a+c+3\eta) \cos b
\nonumber \\
       &+& \cos \beta \cos \theta \sin b \sin(a+\gamma+\alpha-c-3\eta)
\eea
\bea
R_{57} &=& -\sin(\alpha-\gamma) \sin \beta \cos 2\theta \sin(a+c+3\eta) \cos b
\nonumber \\
       &-& \cos(\alpha-\gamma) \sin \beta \cos(a+c+3\eta) \cos b
\nonumber \\
       &-& \cos \beta \cos \theta \sin b \cos(a+\gamma+\alpha-c-3\eta)
\eea
\beq
R_{58} = \sqrt{3} \sin 2\theta \sin(a+c+3\eta) \cos b
\eeq

\bea
R_{61} &=& \cos 2\alpha \cos 2\beta \sin \theta \cos b \cos(a+c-2\gamma-3\eta)
\nonumber \\
       &-& \cos 2\alpha \sin 2\beta \sin 2\theta \cos(a-c+3\eta) \sin b
\nonumber \\
       &+& \sin 2\alpha \sin \theta \cos b \sin(a+c-2\gamma-3\eta)
\eea
\bea
R_{62} &=& -\sin 2\alpha \cos 2\beta \sin \theta \cos b \sin(a+c-2\gamma-3\eta)
\nonumber \\
       &+& \sin 2\alpha \sin 2\beta \sin 2\theta \cos(a-c+3\eta) \sin b
\nonumber \\
       &+& \cos 2\alpha \sin \theta \cos b \sin(a+c-2\gamma-3\eta)
\eea
\bea
R_{63} &=& -\sin 2\beta \sin \theta \cos b \cos(a+c-2\gamma-3\eta)
\nonumber \\
       &-& \cos 2\beta \sin 2\theta \sin b \cos(a-c+3\eta)
\eea
\bea
R_{64} &=& \cos(\alpha+\gamma) \cos \beta \cos 2\theta \cos(a-c+3\eta) \sin b
\nonumber \\
       &-& \sin(\alpha+\gamma) \cos \beta \sin(a-c+3\eta) \sin b
\nonumber \\
       &+& \sin \beta \cos \theta \cos b \cos(a+c+\gamma-\alpha-3\eta)
\eea
\bea
R_{65} &=& -\sin(\alpha+\gamma) \cos \beta \cos 2\theta \cos(a-c+3\eta) \sin b
\nonumber \\
       &-& \cos(\alpha+\gamma) \cos \beta \sin(a-c+3\eta) \sin b
\nonumber \\
       &+& \sin \beta \cos \theta \cos b \sin(a+c+\gamma-\alpha-3\eta)
\eea
\bea
R_{66} &=& -\cos(\alpha-\gamma) \sin \beta \cos 2\theta \cos(a-c+3\eta) \sin b
\nonumber \\
       &-& \sin(\alpha-\gamma) \sin \beta \sin(a-c+3\eta) \sin b
\nonumber \\
       &+& \cos \beta \cos \theta \cos b \cos(a+c+\gamma+\alpha-3\eta)
\eea
\bea
R_{67} &=& -\sin(\alpha-\gamma) \sin \beta \cos 2\theta \cos(a-c+3\eta) \sin b
\nonumber \\
       &+& \cos(\alpha-\gamma) \sin \beta \sin(a-c+3\eta) \sin b
\nonumber \\
       &+& \cos \beta \cos \theta \cos b \sin(a+c+\gamma+\alpha-3\eta)
\eea
\beq
R_{68} = -\sqrt{3} \sin 2\theta \cos(a-c+3\eta) \sin b
\eeq

\bea
R_{71} &=& -\cos 2\alpha \cos 2\beta \sin \theta \cos b \sin(a+c-2\gamma-3\eta)
\nonumber \\
       &-& \cos 2\alpha \sin 2\beta \sin 2\theta \sin(a-c+3\eta) \sin b
\nonumber \\
       &+& \sin 2\alpha \sin \theta \cos b \cos(a+c-2\gamma-3\eta)
\eea
\bea
R_{72} &=& \sin 2\alpha \cos 2\beta \sin \theta \cos b \sin(a+c-2\gamma-3\eta)
\nonumber \\
       &+& \sin 2\alpha \sin 2\beta \sin 2\theta \sin(a-c+3\eta) \sin b
\nonumber \\
       &+& \cos 2\alpha \sin \theta \cos b \cos(a+c-2\gamma-3\eta)
\eea
\bea
R_{73} &=& -\sin 2\beta \sin \theta \cos b \sin(a+c-2\gamma-3\eta)
\nonumber \\
       &+& \cos 2\beta \sin 2\theta \sin b \sin(a-c+3\eta)
\eea
\bea
R_{74} &=& \cos(\alpha+\gamma) \cos \beta \cos 2\theta \sin(a-c+3\eta) \sin b
\nonumber \\
       &+& \sin(\alpha+\gamma) \cos \beta \cos(a-c+3\eta) \sin b
\nonumber \\
       &-& \sin \beta \cos \theta \cos b \sin(a+c+\gamma-\alpha-3\eta)
\eea
\bea
R_{75} &=& -\sin(\alpha+\gamma) \cos \beta \cos 2\theta \sin(a-c+3\eta) \sin b
\nonumber \\
       &+& \cos(\alpha+\gamma) \cos \beta \cos(a-c+3\eta) \sin b
\nonumber \\
       &+& \sin \beta \cos \theta \cos b \cos(a+c+\gamma-\alpha-3\eta)
\eea
\bea
R_{76} &=& -\cos(\alpha-\gamma) \sin \beta \cos 2\theta \sin(a-c+3\eta) \sin b
\nonumber \\
       &+& \sin(\alpha-\gamma) \sin \beta \cos(a-c+3\eta) \sin b
\nonumber \\
       &-& \cos \beta \cos \theta \cos b \sin(a+c+\gamma+\alpha-3\eta)
\eea
\bea
R_{77} &=& -\sin(\alpha-\gamma) \sin \beta \cos 2\theta \sin(a-c+3\eta) \sin b
\nonumber \\
       &-& \cos(\alpha-\gamma) \sin \beta \cos(a-c+3\eta) \sin b
\nonumber \\
       &+& \cos \beta \cos \theta \cos b \cos(a+c+\gamma+\alpha-3\eta)
\eea
\beq
R_{78} = \sqrt{3} \sin 2\theta \sin(a-c+3\eta) \sin b
\eeq

\beq
R_{81} = \frac{\sqrt{3}}{2} \cos 2\alpha \sin 2\beta \sin^2 \theta
\eeq
\beq
R_{82} = -\frac{\sqrt{3}}{2} \sin 2\alpha \sin 2\beta \sin^2 \theta
\eeq
\beq
R_{83} = -\frac{\sqrt{3}}{2} \cos 2\beta \sin^2 \theta
\eeq
\beq
R_{84} = -\frac{\sqrt{3}}{2} \cos(\alpha+\gamma) \cos \beta \sin 2\theta
\eeq
\beq
R_{85} = \frac{\sqrt{3}}{2} \sin(\alpha+\gamma) \cos \beta \sin 2\theta
\eeq
\beq
R_{86} = \frac{\sqrt{3}}{2} \cos(\alpha-\gamma) \sin \beta \sin 2\theta
\eeq
\beq
R_{87} = \frac{\sqrt{3}}{2} \sin(\alpha-\gamma) \sin \beta \sin 2\theta
\eeq
\beq
R_{88} = 1-\frac{3}{2}\sin^2 \theta
\eeq

Recall that $\eta \equiv \phi/\sqrt{3}$.

\section{Summary/Conclusions}

It has been shown that the operators from reference \cite{me} provide a means for finding the irreps of $SU(3)$ by the construction of the ladder operators.  The two fundamental reps and the octet rep have been exhibited explicitly.  The highest weight state for any representation was found thus enabling the calculation of any state within any irrep.  A determination of the ranges of the angles in the Euler angle parameterization was made and the calculation of WCG coeffients was discussed.  Therefore a more complete description of the group $SU(3)$, its manifold and its explicitly parameterized irreps has been given than has been done in the past.

The Clebsch-Gordon coefficients (or WCG coefficients) were calculated by J.J. de Swart in \cite{deSwart} using only algebraic properties.  The operators given here could mimic those results as well.  The Euler angle parameterization was given by Beg and Ruegg along with a calculation of the differential operators that are valid for some particular cases (namely those for which there exists an isosinglet) \cite{BandR}.  Holland \cite{H} and Nelson \cite{N} originally gave an accounts of the irreps of $SU(3)$, but the rep matrices were presented in a somewhat less managable form.  These were also investigated by Akyeampong and Rashid \cite{AandR}.  It is anticipated that this more managable account will lead to new applications.  It has already proven to be useful in the description of three state systems.  This will be discussed elsewhere.

\section{Acknowledgments} 

One of us (M.S.B.) would like to thank Prof. L. C. Biedenharn (now deceased).  I could not give him too much credit here.  I would also like to thank Prof. Duane Dicus whose help and support enabled the completion of this paper.  This research was supported in part by the U.S. Department of Energy under Contract No. DE-EG013-93ER40757.

\end{document}